\documentclass{elsart3p}

\usepackage{graphics}
\usepackage{epsfig}

\usepackage{amssymb}

\begin{document}

\begin{frontmatter}



\title{Ways to constrain the away side jet in Au + Au collisions in PHENIX}


\author{Jiangyong Jia for the PHENIX Collaboration}

\address{Columbia University and Nevis Laboratories, Irvington, NY 10533, USA}
\address{Current address: State University of New York at Stony Brook, Stony Brook, NY 11790, USA}

\begin{abstract}
We discussed methods used by the PHENIX to constrain the flow
background in the two particle jet correlation. Both the
background level and elliptic flow can be reliably decomposed from
the jet contribution. We also studied the non-flow contribution to
the reaction plane elliptic flow due to dijets. We found the jet
bias is negligible in PHENIX, when the reaction plane is measured
at Beam Beam Counter acceptance ($3<|\eta|<4$).
\end{abstract}


\end{frontmatter}

\section{Introduction}

Two particle azimuth correlation is a useful tool to study the
strongly interacting partonic matter believed to has been created
at RHIC Au+Au collisions. The typical correlation function can be
decomposed into a jet part $J(\Delta\phi)$ and an underlying event
part $\xi$ that is modulated by the elliptic
flow~\cite{Adler:2005ee}:
\begin{eqnarray}
\label{eq:1} C(\Delta\phi) = J(\Delta\phi) + \xi
(1+2v_2^tv_2^a\cos2\Delta\phi)
\end{eqnarray}
In the case that dijet is not modified by the medium,
$J(\Delta\phi)$ behaves just like in p+p collisions.  The RHIC
data has shown a rich modification pattern, which is dependent on
the $p_T$ of both particles and can be characterized into four
qualitatively distinct regions. 1) A broadening jet shape at the
away side and enhancement of the jet multiplicity at both the near
and away side at low $p_T$~\cite{Adams:2005ph}; 2) A flat or
possible volcano-shaped away side jet pairs at intermediate
$p_T$~\cite{Adler:2005ee}; 3) A seemly complete disappearance of
away side jet or equivalently a flat away side extended to the
near side at moderately high $p_T$~\cite{Adler:2002tq};4)
Reappearance of the away side jet peak at very high
$p_T$\cite{Adams:2006yt}. Several models were proposed to explain
various aspects of the modification pattern, but so far no model
can consistently describe all four regions on a quantitative
level.

The medium effects on the dijets can be quantified by
$I_{\rm{AA}}$, which is the ratio of the jet yield per trigger in
A+A collisions to that in p+p collisions. Since the modification
is always on the jet pairs, the following relation holds between
the per-trigger yield using the high $p_T$ particles as triggers
(type a) and that using low $p_T$ particles as triggers (type b):
\begin{eqnarray}
\label{eq:eq1} R_{\rm{AA}}^aI_{\rm{AA}}^a   =
R_{\rm{AA}}^bI_{\rm{AA}}^b &=& \frac{{\rm{JetPairs}_{\rm{AA}}
}}{{N_{\rm{coll}} \times \rm{JetPairs}_{\rm{pp}} }}
\end{eqnarray}
where $R_{\rm{AA}}$ is the single particle suppression factor,
JetPairs$_{\rm{AA}}$ and JetPairs$_{\rm{pp}}$ represent the
average number of jet pairs in one A+A collision and one p+p
collision, respectively. One would expect
$I_{\rm{AA}}<R_{\rm{AA}}$ for hadron-hadron correlation due to
trigger surface bias. In contrast, the medium is transparent to
the leading photons in direct photon-jet correlations. One expect
that the away side jet behave exactly as the single jet
suppression: $I_{\rm{AA}} = R_{\rm{AA}}$.

The level of accuracy in extracting the jet signal depends on how
well one determines the elliptic flow of both particles, $v_2^t$,
$v_2^a$ and combinatoric background level $\xi$. PHENIX measures
the elliptic flow through the reaction plane (RP) method, where
the event plane (EP) is determined by the BBC at forward region
($3<|\eta|<4$). The systematic error on the $v_2$ is dominated by
the EP resolution, $\delta v_2^t /v_2^t = \delta v_2^a
/v_2^a=\epsilon_{\rm{reso}}$. $\xi$ represents the ratio of the
combinatoric background in same event to that in mixed event:
$\left<n_{t}n_{a}\right> = \xi
\left<n_{t}\right>\left<n_{a}\right>$. It is typically slightly
above 1 due to finite multiplicity fluctuation in a typical
centrality bin. The uncertainty on the correlation function can be
expressed as:
\begin{eqnarray}
\label{eq:3} \nonumber\delta C &=& 2\left(\xi\delta v_2^t v_2^a +
\xi\delta v_2^a v_2^t + \delta\xi v_2^t v_2^a\right)\cos
2\Delta\phi + \delta\xi
\\ &\approx& \left(2\epsilon_{\rm{reso}}\xi+\delta\xi\right)2 v_2^t v_2^a\cos
2\Delta\phi + \delta\xi
\end{eqnarray}
Given the typical values of various factors: $\xi\approx 1$,
$\delta\xi <0.01$, $2v_2^t v_2^a < 0.1$, and
$\epsilon_{\rm{reso}}\approx 0.05-0.1$, the above formula can be
simplified as
\begin{eqnarray}
\label{eq:4} \delta C \approx 4\epsilon_{\rm{reso}} v_2^t
v_2^a\cos2\Delta\phi + \delta\xi
\end{eqnarray}
\section{Constrain Background Level $\xi$}
PHENIX has previously used two methods in determining the $\xi$:
the absolute normalization (ABN) method~\cite{Adler:2004zd} and
Zero Yield At Minimum (ZYAM)
method~\cite{Adler:2005ee,Ajitanand:2005jj}. The ZYAM method
assumes there is a point in $\Delta\phi$ where jet yield is zero,
i.e the background term ``kisses'' the correlation function. The
$\xi$ value from ZYAM method is a upper limit since the background
can't be higher than the total in Eq.\ref{eq:1}. The ABN method
calculate $\xi$ directly by assuming a certain relation between
multiplicity and centrality.

\begin{figure}[ht]
\epsfig{file=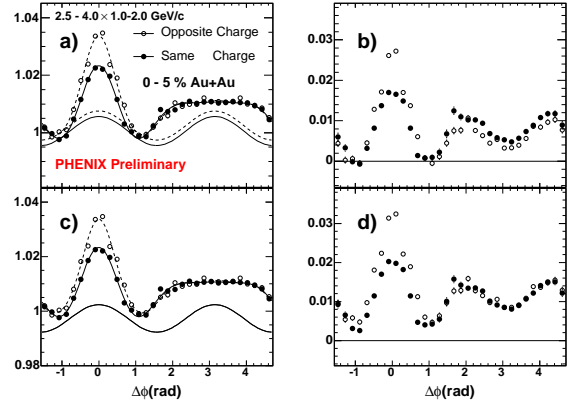,width=1.0\columnwidth}
\caption{\label{fig:ch1} a),c) The correlation function in 0-5\%
centrality bin for opposite-sign pairs (open circles) and
same-sign charged pairs (closed circles). The lines indicate the
background level determined via ZYAM method (a)) and
charge-dependent method (c)). b),d) The background subtracted
distribution for using ZYAM (b)) and charge-dependent method
(d)).}
\end{figure}

Here we discuss a third method in constraining $\xi$, based on the
combined information from opposite-sign charged pairs (OSC) and
same-sign charged pairs (SSC). Fig.\ref{fig:ch1}a shows the
azimuthal distributions for OSC and SSC for top 0-5\% Au+Au
collisions for $2.5<p_{T,\rm{trig}}<4$ GeV/$c$ and
$1<p_{T,\rm{asso}}<2$ GeV/$c$. The OSC correlation at the near
side is larger than that for SSC, while the charge combination has
no effect on the back-to-back correlations. The difference of the
jet strength at the near side is a consequence of the charge
ordering effect in the jet fragmentation process, which leads to
more OSC pairs than SSC pairs. One would not expect any charge
correlation between the pair at the $\Delta\phi\approx\pi$, since
they come from different jets. The ZYAM method clearly finds
different $\xi$ values between the two although their true values
should be identical. Fig.\ref{fig:ch1}b shows the jet signal
estimated independently for OSC and SSC using the ZYAM approach.
The estimated away side jet yield for OSC and SSC pairs are
clearly different due to the difference in $\xi$ from ZYAM.

In both cases, the ZYAM minimum is at around the same
$\Delta\phi_0\approx0.8$. The difference at that point indicated
different amount of near side jet contribution. Given that the
near side jet width are the same between SSC and OSC
pairs~\cite{Kravitz:2005wc}, jet contribution can be written
explicitly for the two cases as,
\begin{eqnarray}
J^{\rm{S}} &=& J^{S}_N(\Delta\phi)+J_A(\Delta\phi)\\\nonumber
J^{\rm{O}} &=& J^{\rm{O}}_N(\Delta\phi)+J_A(\Delta\phi)=
A_0J^{\rm{S}}_N(\Delta\phi)+J_A(\Delta\phi)
\end{eqnarray}
Define $\Delta\xi = (A_0-1)J^{\rm{SSC}}_N(\Delta\phi_0)$, i.e. the
difference in the $\xi$ value between OSC and SSC pairs, we
got
\begin{eqnarray} \nonumber J^{\rm{S}}_N(\Delta\phi_0) &=&
\frac{\Delta\xi}{A_0-1}
=\frac{\Delta\xi}{J^{\rm{O}}_N(0)/J^{\rm{S}}_N(0)-1}\\
&=&\Delta\xi\frac{J^{\rm{S}}_N(0)}{J^{\rm{O}}_N(0)-J^{\rm{S}}_N(0)}\\\nonumber\\
J^{\rm{O}}_N(\Delta\phi_0)
&=&\Delta\xi\frac{J^{\rm{O}}_N(0)}{J^{\rm{O}}_N(0)-J^{\rm{S}}_N(0)}
\end{eqnarray}
The near side jet contribution at the $\Delta\phi_0$ can be
calculated from the near side jet peak value (at $\Delta\phi=0$),
which is given by ZYAM method. Since the jet fraction is large at
the peak region, fractional error of the peak values due to
uncertain on $\xi$ from ZYAM is small. We include it in final
systematic error. The results of this procedure is shown in
Fig.\ref{fig:ch1}d. The subtracted distribution are identical on
the away side but shifted above 0 as expected.

The charged dependence method is useful when the statistic is
good, such that $\xi$ is well constrained at the ZYAM minimum and
$J_N(\Delta\phi_0)$ can be reliability estimated. We recognize
that there could be contributions from away side jet at ZYAM
minimum also, $J_A(\Delta\phi_0)$, which is inaccessible in
current approach. Thus the $\xi$ value obtained in this method
could still be too big. Table.\ref{tab:xi} summarize the $\xi$
values obtained from the three methods.

\begin{table}
\begin{center}
\caption{\label{tab:xi} $\xi$ values calculated for
$2.5<p_{T,\rm{trig}}<4.0$ GeV/$c$ and $1.0<p_{T,\rm{asso}}<2.0$
GeV/$c$ bin compared between three different normalization method
for several centrality selections. }
\end{center}
\begin{tabular}{|c|c|c|c|}\hline
 Centrality  & ZYAM               & absolute & charge dependent\\\hline
 0-5\%       & $1.0018\pm0.0004$  &$1.0023\pm0.0002$& $0.998\pm0.002$\\
20-30\%      & $1.015\pm0.0015$  &$1.012\pm0.003$&$1.004\pm0.006$\\
50-60\%      & $1.076\pm0.009$  &$1.07\pm0.02$
&$1.054\pm0.009$\\\hline
\end{tabular}
\end{table}



\section{Constrain $v_2$ from Reaction-Plane Dependent Correlation}
\begin{figure}[ht]
\begin{center}
\epsfig{file=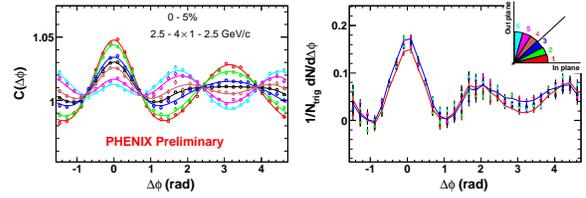,width=1.0\columnwidth}
\caption{\label{fig:rp} a) Correlation function for various 6
trigger direction bin and the trigger integrated bin (the center
curve. b) The background subtracted per-trigger yields, the insert
figure shows the 6 trigger bins.}
\end{center}
\end{figure}

Let's define $\sigma_n = \langle\cos
n(\Psi_{\rm{EP}}-\Psi_{\rm{RP}})\rangle$ as EP resolution for
$n$th order harmonics using the elliptic flow RP. According
to~\cite{Bielcikova:2003ku}, the pair distribution when the
triggers are selected in a limited angular bite of $\phi_s\pm c$
with respect to reaction plane is
\begin{eqnarray}
\label{eq:rp} &&C_{c,\phi_s} = J_{c,\phi_s}(\Delta\phi)+\xi(1 +
2v_2^a\frac{b}{a} \cos2\Delta\phi)\\\nonumber&&
\begin{array}{l}
 \left\{ \begin{array}{l}
a = 1 + 2v_2^t \cos 2\phi_s \frac{{\sin 2c}}{{2c}}\sigma_2
\\
b = v_2^t  + \cos 2\phi_s \frac{{\sin 2c}}{{2c}}\sigma_2  + v_2^t
\cos 4\phi_s \frac{{\sin4c}}{{4c}}\sigma_4
 \end{array} \right. \\
  \\
 \end{array}
\end{eqnarray}
Fig.\ref{fig:rp}a shows the CFs in 0-5\% central Au+Au collisions
for 6 angular bins in $15^o$ steps. Fig.\ref{fig:rp}b shows the
jet yield after subtracting the flow terms calculated according to
Eq.\ref{eq:rp}. Although the CFs change dramatically from in plane
to out of plane, the calculated flow term tracks the true flow
background nicely. Given the small eccentricity in 0-5\%, we can
safely assume that the jet yield, $J_{c,\phi_s}$ does not depends
on the trigger direction. In this case, the small difference
between the jet functions in Fig.\ref{fig:rp}b can be used to
further constrain the $v_2$.

\section{Non-flow Effects from Jets}
Reliable extraction of the jet signal requires accurate
determination of $v_2^t$ and $v_2^a$. To this end, contributions
from non-flow correlations that lead to azimuth correlations not
related to the true RP direction, need to be studied. The non-flow
correlations include transverse momentum conservation effects,
resonance decays, HBT correlations that are important at low
$p_T$~\cite{Borghini:2000cm,Borghini:2001vi} and jet correlations
that are important at high $p_T$~\cite{Adams:2004wz}. The non-flow
correlations affect the $v_2$ values either by changing the EP
resolution or cause fake $v_2$ by biasing the EP direction towards
the now-flow particles. The former is small if the non-flow
particle multiplicity is small. The later can be suppressed in
PHENIX if the correlations are limited in a narrow $\Delta\eta$
window, such as the resonance and intra-jet correlations, thanks
to the large $\eta$ separation between BBC and central arm.
However, the inter-jet correlations can still bias the BBC
reaction plane determination, due to their broad distribution in
$\Delta\eta$.

\begin{figure}[ht]
\begin{center}
\epsfig{file=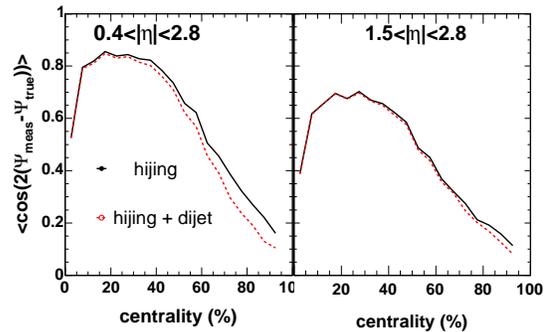,width=0.9\columnwidth}
\caption{\label{fig:v2reso} The effect of the dijet on reaction
plane resolution as function of centrality in different rapidity
windows. The embedded jet is at mid-rapidity.}
\end{center}
\end{figure}

We study the biases due to dijet correlations by embedding dijet
pairs into background events with realistic flow modulation. The
background Au+Au events are simulated with HIJING, which was
checked to reproduce the charged hadron multiplicity in $\eta$
from PHOBOS. Elliptic flow is implemented by applying a track by
track weight for each HIJING event:
\[
w(b,p_T,\eta)=1+2v_2(b,p_T,\eta)\cos2(\phi-\Psi)
\]
where the $\Psi$ is the direction of the impact parameter $b$. The
centrality, $p_T$ dependence of the $v_2$ is tuned according to
the PHENIX measurement~\cite{Adare:2006ti}. We used a common
$\eta$ dependence from PHOBOS~\cite{Manly:2005zy} for all
centrality selections. The dijet pairs are generated from PYTHIA
event generator, requiring a leading particle above 6 GeV/$c$ at
mid-rapidity ($|\eta|<0.35$). This corresponds to typical energy
of $6/\langle z \rangle\approx 10$ GeV/$c$ for the dijets.

\begin{figure}[ht]
\begin{center}
\epsfig{file=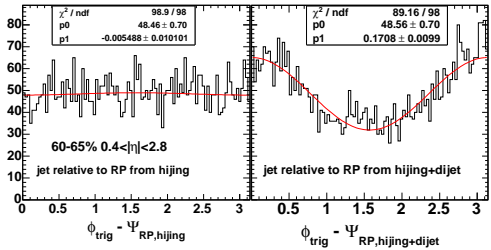,width=0.9\columnwidth}
\caption{\label{fig:rpembed} The distribution of the leading
particle from the dijets relative to the event plane calculated
from HIJING only (left) and event plane from the embedded event
(right).}
\end{center}
\end{figure}

The effects of the dijets are evaluated by comparing the event
plane before and after the embedding. Since dijet pairs are random
with respect to the RP, the EP resolution for combined event is
worse as shown in Fig.\ref{fig:v2reso}. The difference in EP
resolution is small except in peripheral bins where the event
multiplicity is small. And the difference become negligible in
more forward $\eta$ window. On the other hand, dijet tends to pull
the event plane towards the dijet direction, resulting in a fake
$v_2$ for the leading hadrons as shown in Fig.\ref{fig:rpembed},
where the relative azimuth distribution between the leading
hadrons and EP from either the HIJING event or the combined event
are plotted. The dijets are clearly correlated with the EP
determined from the combined event, thus have a fake $v_2$.

\begin{figure}[ht]
\begin{center}
\epsfig{file=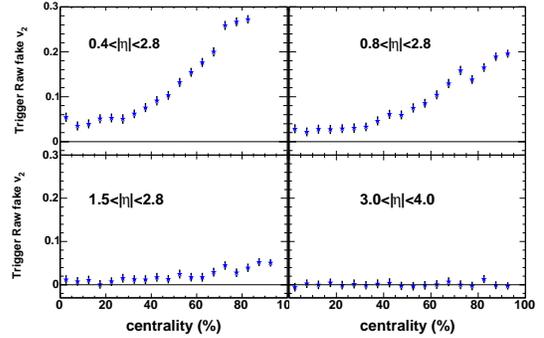,width=0.9\columnwidth}
\caption{\label{fig:v2bias} The fake $v_2$ of the leading particle
as function of centrality using the EP determined in four $\eta$
windows. The embedded jet is at mid-rapidity.}
\end{center}
\end{figure}

The size of the fake $v_2$ depends on the rapidity gap between the
triggering hadron and the subevent used to determine the event
plane. Due to the away side jet swing, one would expect the this
bias persists to large $\eta$ region. Fig.\ref{fig:v2bias} shows
the centrality dependence of fake $v_2$ for various rapidity
window. This fake $v_2$ is the raw signal extracted from fit in
Fig.\ref{fig:rpembed} without divided by EP resolution. The fake
$v_2$ decreases as the subevent moves towards large $\eta$. When
the subevent is $3<|\eta|<4$ (BBC acceptance), the fake $v_2$
becomes almost negligible.

\begin{figure}[ht]
\begin{center}
\epsfig{file=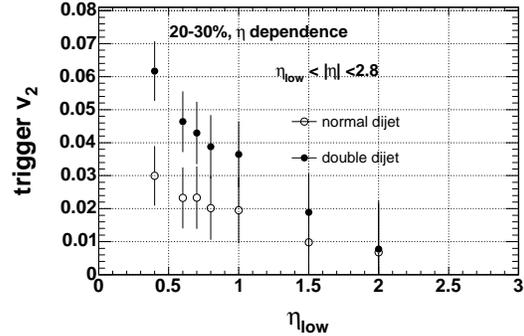,width=0.9\columnwidth}
\caption{\label{fig:v2biasw} The EP resolution corrected fake
$v_2$ as function of $\eta_{\rm{low}}<|\eta|<2.8$ for normal dijet
and enhanced dijet.}
\end{center}
\end{figure}

So far, we have assumed normal dijet in our study. RHIC data
indicates that there is a broadening of the away side jet and an
increase of jet multiplicity at low $p_T$ by about factor of
2~\cite{Adams:2005ph,Jia:2005av}. To account for that, we increase
the PYTHIA dijet multiplicity by factor of 2 and redo the study of
the rapidity dependence of the fake $v_2$. We did not take into
account the broadening of the near side jet in $\eta$, since it is
dominated by the away side jet swing. Fig.\ref{fig:v2biasw} shows
the fake $v_2$ corrected for the EP resolution as function of
rapidity window with enhanced dijet and normal dijet. The elliptic
flow is higher when the jet multiplicity is doubled, but it is
still negligible in the BBC acceptance.

The $v_2$ measurements can also be affected by the event by event
fluctuation of $v_2$
~\cite{Miller:2003kd,Zhu:2005qa,Manly:2005zy,Bhalerao:2006tp},
which is a direct consequence to the event by event fluctuation of
the collision geometry. However, since the two particle
correlation method is used in this analysis, the event by event
fluctuations would also contribute to the $v_2$ correlations. In
this sense the $v_2\{2\}$ should be the one to use in
Eq.\ref{eq:1}. Since $v_2\{2\}$ was shown to be consistent with
the BBC RP $v_2$~\cite{Adler:2004cj} up to 4 GeV/$c$ in $p_T$, it
is reasonable to using BBC RP $v_2$ in Eq.\ref{eq:1}.

\end{document}